\documentclass[%
 reprint,
superscriptaddress,
onecolumn,
bibnotes,
 amsmath,amssymb,
]{revtex4-2}

\usepackage{graphicx}
\usepackage{dcolumn}
\usepackage{bm}
\usepackage{xcolor}
\usepackage{float}

\begin{document}

\title{Rapid multiplex ultrafast nonlinear microscopy for material characterization}

\author{Torben L. Purz}
\affiliation{University of Michigan, Department of Physics, Ann Arbor, Michigan 48109, USA}
\author{Blake T. Hipsley}
\affiliation{University of Michigan, Department of Physics, Ann Arbor, Michigan 48109, USA}
\author{Eric. W. Martin}
\affiliation{MONSTR Sense Technologies, LLC, Ann Arbor, Michigan 48104, USA}
\author{Ronald Ulbricht}
\affiliation{Max Planck Institute for Polymer Research, Ackermannweg 10, 55128 Mainz, Germany}
\author{Steven T. Cundiff}
 \email{cundiff@umich.edu}
\affiliation{University of Michigan, Department of Physics, Ann Arbor, Michigan 48109, USA}



\begin{abstract} 
We demonstrate rapid imaging based on four-wave mixing (FWM) by assessing the quality of advanced materials through measurement of their nonlinear response, exciton dephasing, and exciton lifetimes. We use a WSe\textsubscript{2} monolayer grown by chemical vapor deposition as a canonical example to demonstrate these capabilities. By comparison, we show that extracting material parameters such as FWM intensity, dephasing times, excited state lifetimes, and distribution of dark/localized states allows for a more accurate assessment of the quality of a sample than current prevalent techniques, including white light microscopy and linear micro-reflectance spectroscopy. We further discuss future improvements of the ultrafast FWM techniques by modeling the robustness of exponential decay fits to different spacing of the sampling points. Employing ultrafast nonlinear imaging in real-time at room temperature bears the potential for rapid in-situ sample characterization of advanced materials and beyond.
\end{abstract}

\maketitle

\section{Introduction}

Since the invention of the microscope over 400 years ago, the need to improve imaging techniques and modalities for obtaining previously inaccessible information, and obtaining it faster, has been a common theme surrounding microscopy. As new groups of advanced materials shift into focus, the need to image these materials for fundamental science and to characterize them in a manufacturing/fabrication setting has spurred numerous experimental innovations. These materials include two-dimensional quantum materials such as transition metal dichalcogenides (TMDs) and graphene \cite{TMD_Lego,TMD_Photodiode,Photovoltaics,TMD_Laser,TMDQuantumLED,Galan_Wse2,CoherentCouplingPurz,Graphene_defects,Graphene_industry}, III-V semiconductors such as Gallium Arsenide and Gallium Nitride in both science \cite{III-V-APD,III-V-IRPD,III-V-SolarHydrogen} and industry \cite{GaN_industry}, and silicon carbide for electric vehicles \cite{SiC-Forbes}. Material characterization of TMDs and other advanced materials has seen a plethora of techniques, from white-light optical microscopy to photoluminescence imaging \cite{PL_imaging}, micro-reflectance and transmission \cite{MicroReflectance}, scanning tunneling microscopy \cite{Characterization_review}, angle-resolved photoemission spectroscopy \cite{Characterization_review}, Raman spectroscopy \cite{Raman1,Raman2,RamanImaging}, atomic force microscopy imaging \cite{AFM_review}, tip-enhanced spectroscopy \cite{Tip_Markus}, ultrafast nanoscopy \cite{Nanoscopy_Huber}, and four-wave mixing (FWM) imaging \cite{Kasprzak,Kasprzak_Coherent,Kasprzak_LW1,Kasprzak_LW2,JCP}. However, these techniques either convey little information about the material quality \cite{JCP}, or require complicated experimental setups and hours of data acquisition. 

To overcome these limitations, we here introduce rapid multiplex nonlinear imaging, enabled by recent advances in lock-in detection \cite{OPL}, to characterize advanced materials. Specifically, we demonstrate the technique's feasibility on WSe\textsubscript{2} monolayers grown by chemical vapor deposition (CVD), which serve as a canonical example.
We acquire resonant linear reflectance and nonlinear FWM images that provide rich information about the materials quality in real-time while acquiring full exciton dephasing and lifetime maps within a minute. We correlate the findings from these rapid techniques with multi-dimensional coherent imaging spectroscopy (MDCIS) data to corroborate our findings. We distinguish areas of monolayer flakes by their FWM strength and dephasing and decay dynamics, while specifically identifying areas with weak FWM, strong many-body effects, and an increased density of dark states. We show that the dephasing times are radiatively limited by correlating dephasing and lifetime maps. Lastly, we demonstrate how these modalities can be further accelerated in the future by modeling the robustness of decay time fits across a large range of decay times for different sampling point spacing. 
These results constitute an important step toward real-time material characterization of advanced materials on an industrial scale, a necessity for devices to make it out of the lab and into the marketplace.

\section{Experimental methods}

\begin{figure*}[t]
\centering\includegraphics[width=0.7\textwidth]{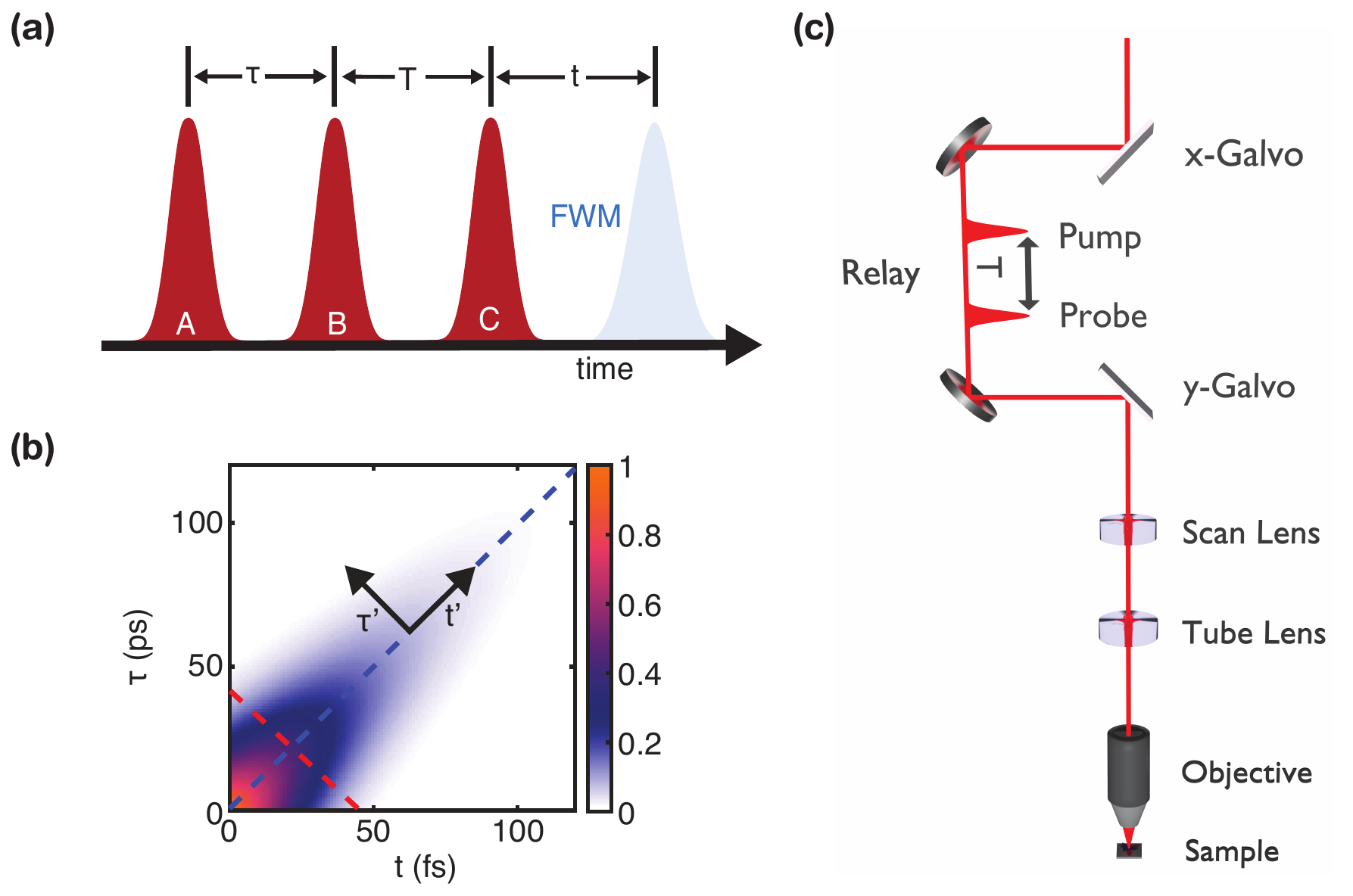}
\caption{\textbf{(a)} Schematic of a three-pulse FWM experiment. The FWM is emitted after the third pulse $C$. By varying different time delays, we can access the dephasing of the exciton coherence and decay of the exciton density.  \textbf{(b)} Temporal evolution of the FWM signal for varying $\tau$ and $t$ delays. Dephasing time and inhomogeneous linewidth can be extracted by taking slices along the axes spanned by the blue and red dashed line. \textbf{(c)} Schematic of the custom-built laser-scanning microscope. The $x$- and $y$-Galvos are relayed onto each other using off-axis parabolic mirrors before being relayed onto the back of the microscope objective with a combination of scan and tube lens.}
\label{fig:Fig1}
\end{figure*}

All modalities of FWM imaging employed in this work are based on the MDCIS setup detailed in previous work \cite{JCP}. In this technique, as visualized in Fig.\,\ref{fig:Fig1}(a), three pulses ($A, B, C$) impinge on the sample.
The three pulses $A, B, C$ jointly generate a FWM signal by each converting the sample state between populations and coherences. As such, scanning different temporal delays between $A$, $B$, and $C$ alters the resulting FWM and gives access to dephasing of the coherences and decay of exciton population, respectively.
After the third interaction ($C$), the emitted FWM signal is heterodyne-detected with the fourth pulse ($D$, not shown). Each of the four pulses is tagged with a unique radio-frequency using acousto-optic modulators \cite{Nardin}, and the interference between the FWM signal emitted from the sample and pulse $D$ is phase-sensitively detected using a custom lock-in amplifier \cite{OPL}.

A complete five-dimensional MDCIS data set, obtained by scanning all three time delays in Fig.\,\ref{fig:Fig1}(a) while acquiring an image for every fixed $\tau$, $t$, and $T$ position, can be valuable for an in-depth physical analysis of the underlying sample system, as demonstrated in \cite{JCP}. However, for material characterization, the need for speed often outweighs the need for full spectroscopic or temporal detail and often more rudimentary temporal or spectral data is sufficient to assess the sample quality. Quantities that can be used to assess the quality of a sample include the FWM strength, the exciton dephasing time, and the exciton population decay time across the sample. All of these modalities can be extracted from complete MDCIS measurements with data acquisition times of at least 30\,minutes, but specific parameter images can be acquired within a minute or less. FWM strength can be assessed by setting all time delays in Fig.\,\ref{fig:Fig1}(a) to zero and recording a single image by scanning the laser beam across the sample. Similarly, decay maps of the sample can be obtained by setting the $\tau$- and $t$-delay to zero while scanning the $T$-delay. Obtaining dephasing time maps involves scanning the $\tau$ and $t$-delay simultaneously while leaving the $T$-delay stationary. This is illustrated in Fig.\,\ref{fig:Fig1}(b), where the temporal response of the sample to varying $\tau$ and $t$ delays is shown. 
According to \cite{Fits}, the FWM signal in the time-domain, assuming $\delta$-function pulses and the Markovian approximation is
\begin{equation}
    R(t',\tau') = R_{0} e^{-(t'/T_2 + i\omega_0 \tau' + \sigma^2 {\tau'}^2/2)} \Theta (t'-\tau') \Theta (t'+\tau')\,.
\end{equation}

\noindent Here, $R_0$ is the amplitude at time zero, $\omega_0$ is the center resonance frequency, $T_2$ is the dephasing time, $\sigma$ is the inhomogeneous linewidth, and the $\Theta$'s are unit step functions ensuring causality. Here we have defined $t'=t+\tau$ as the time coordinate along the photon-echo  (Photon-echo delay), and $\tau'=t-\tau$ as the coordinate orthogonal to the photon echo. The resulting coordinate system is spanned by the blue and red dashed lines in Fig.\,\ref{fig:Fig1}(b). The authors in Refs. \cite{Kasprzak_LW1,Kasprzak_LW2} fit slices along the diagonal (blue dashed line) and cross-diagonal (red dashed line) in the time-domain to extract dephasing times and inhomogeneous linewidths, but still acquire a full, two-dimensional time-domain signal. Here, instead of scanning the full range of $\tau$- and $t$-delays, we propose scanning along the diagonal of the photon-echo, yielding the dephasing time $T_2$ by fitting the data with a uni-exponential decay. 
This measurement can be obtained in less than a minute and thus yields rapid information about the dephasing time across the sample.

Rapid scanning while isolating a single frequency modulated signal in the presence of several frequency-modulated signals requires a custom lock-in amplifier for efficient suppression of extraneous modulations \cite{OPL}. Here, we use a pixel dwell time of 240\,$\mu$s for all imaging measurements, which is currently limited by the modulation frequencies of the FWM and does not constitute a fundamental limit of this technique. Because the FWM signal is phase-resolved, we can further employ coherent averaging for each image to improve the signal-to-noise ratio (SNR).

To obtain images we use a custom-built laser scanning microscope whose design schematic is shown in Fig.\,\ref{fig:Fig1}(c). We employ separate $x$- and $y$-galvo mirrors, which are relayed onto each other using off-axis parabolic mirrors. Subsequently, broadband, large field-of-view scan and tube lenses are used to image the galvo mirrors onto the microscope objective (Nikon 20x, NA=0.4). In Fig.\,\ref{fig:Fig1}(c), we show the pulse configuration for acquiring exciton lifetime maps with pulses $A$ and $B$ ("Pump") and $C$ ("Probe") and $D$, respectively, being overlapped in time.

Many materials show sub-picosecond temporal dynamics and broadband spectral features \cite{CoherentCouplingPurz,JCP,MDCS_CT,Graphene}, wherefore transform-limited, broadband laser pulses are required. However, depending on the sample of interest, the center wavelength of the laser needs to be easily adjusted. Combining high wavelength tunability with in-situ pulse compression requires using a spatial light modulator (SLM)-based pulse shaper.
In our experiment, we spatially disperse the laser pulses using a 1200 grooves/mm grating and subsequently focus onto the SLM (Meadowlark Optics 1x12K Linear SLM) with a concave 10\,cm cylindrical mirror. We generate a phase mask on the SLM to specifically correct for the dispersion acquired from the optical elements in the experimental setup. We employ phase-resolved cross-correlation with a known reference pulse, characterized by SHG-FROG, to characterize the resulting pulses at the sample. Applying the appropriate phase mask compresses the pulses to 30\,fs (full width half maximum), which is near the transform limit of 24\,fs. Figures showing the pulse in both time and frequency domains that highlight the capabilities of the SLM pulse shaper can be found in the Supplemental Material \cite{Supplement}.

\section{Results}

\begin{figure*}[b]
\centering\includegraphics[width=1\textwidth]{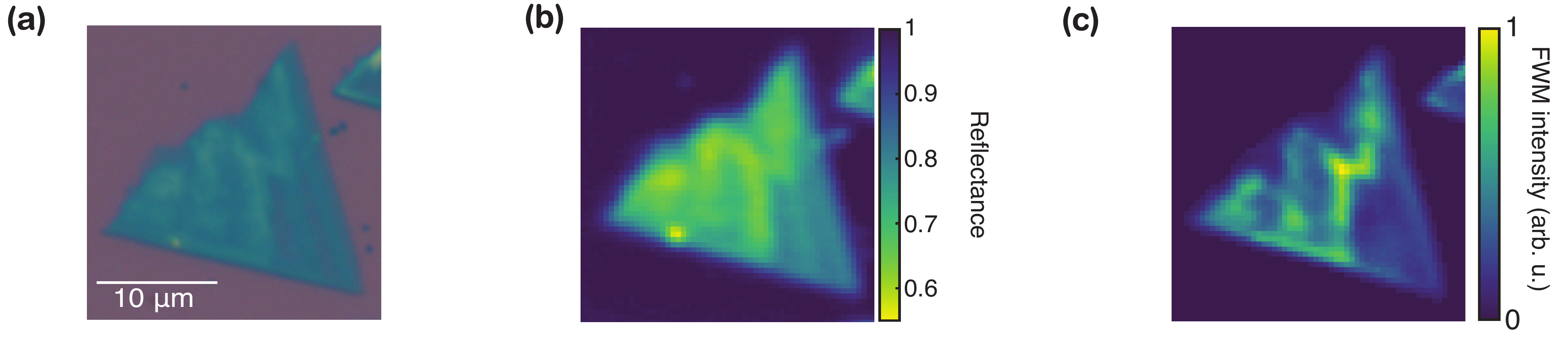}
\caption{\textbf{(a)} White-light microscopy image (false color) of a CVD-grown WSe\textsubscript{2} monolayer. \textbf{(b)} Resonant integrated reflectance from 1600\,meV to 1700\,meV of the WSe\textsubscript{2} monolayer. Here, we set the substrate to have a reflectance of one. \textbf{(c)} FWM intensity image of the WSe\textsubscript{2} monolayer.}
\label{fig:Fig2}
\end{figure*}

An image of the sample acquired with a conventional white-light microscope is shown in Fig.\,\ref{fig:Fig2}(a). The sample is a commercially available CVD-grown flake of WSe\textsubscript{2} (6Carbon Technologies), grown on a separate substrate and transferred onto a new substrate of SiO\textsubscript{2}/Si.
The monolayer shows an uneven structure due to residue remaining from the transfer process, a common problem in CVD grown materials \cite{CVD_transfer}. A resonant (with the exciton) integrated reflectance image is shown in Fig.\,\ref{fig:Fig2}(b). Here, we use a sample point on the substrate to reference a reflectance of one and integrate over the laser spectrum spanning a range from 1600\,meV to 1700\,meV. The sample reflectance is influenced by both reflections from the sample and back-reflected signal from the substrate that is absorbed in the sample. The spatial structure of the integrated reflectance coincides with the spatial structure visible in the white light microscopy image in Fig.\,\ref{fig:Fig2}(a). Nonetheless, differences (e.g., at the bottom of the sample) remain.

A FWM intensity image of the sample is shown in Fig.\,\ref{fig:Fig2}(c). It indicates a strong spatial dependence of the FWM strength, which has been attributed to local strain profiles, changes in the dielectric environment, doping, trapped charges, impurities, defect densities, and distribution of dark states \cite{Kasprzak_LW2,Kasprzak_LW1,Kasprzak_Coherent,ChernikovDielectric,JCP}.
Some regions of stronger FWM correlate with areas of weaker reflectance (e.g., the bright center structure). In contrast, some areas of weaker reflectance (e.g., towards the center-left of the sample) show an overall weaker FWM signal. Moreover, some areas of stronger reflectance, such as the bottom of the sample, show a stronger FWM signal. These observations highlight one of the benefits of nonlinear FWM imaging: while white light microscopy and even resonant linear micro-reflectance spectroscopy can be helpful for sample characterization, the sensitivity of FWM to material changes, including doping, defects, strain, dielectric environment, and dark state distribution changes \cite{Kasprzak,ChernikovDielectric,Eric_TMD,JCP}, yields more detailed information about the quality of a sample.

\begin{figure*}[t]
\centering\includegraphics[width=0.7\textwidth]{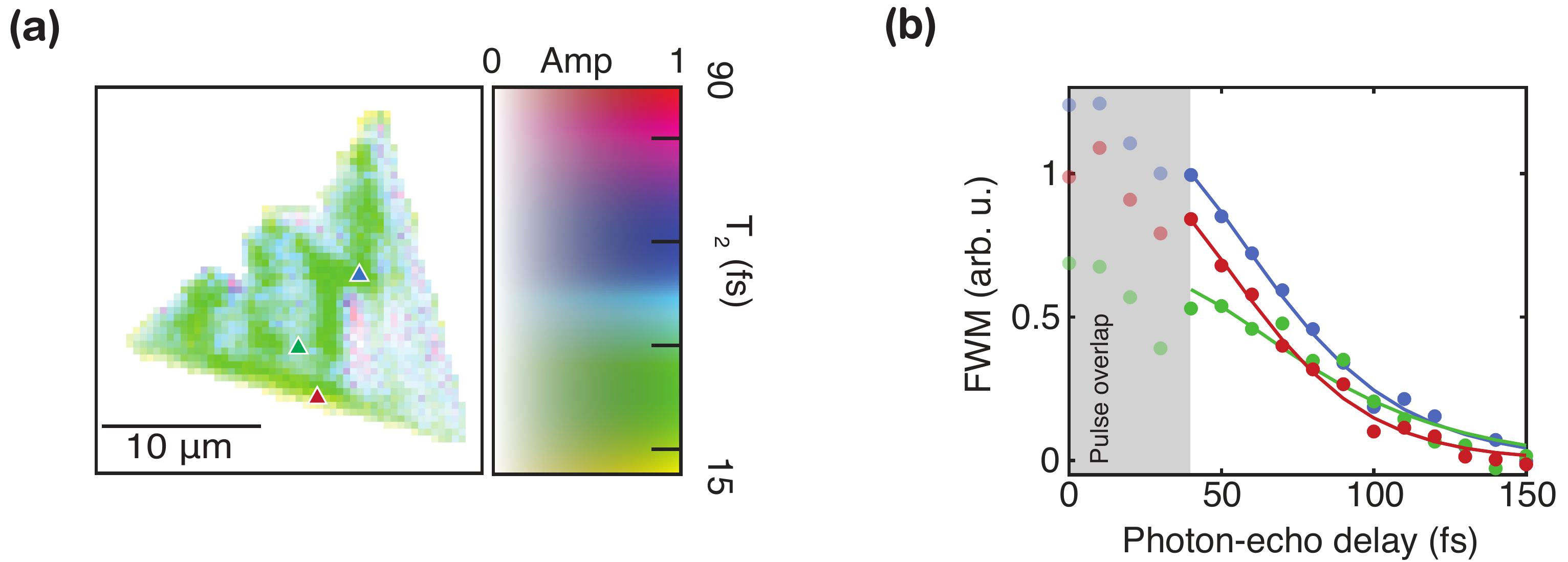}
\caption{\textbf{(a)} Dephasing time $T_2$ map of the sample obtained by fitting a uni-exponential decay in the time-domain. The hue level signifies the dephasing time while the saturation signifies the FWM strength. \textbf{(b)} FWM amplitude dephasing curves at select sample spots marked with colored triangles in (a). Uni-exponential fits are plotted as solid lines. Only sample points with photon echo times $t+\tau = t'\geq40$\,fs are fitted to exclude finite pulse effects present for early $t'$.}
\label{fig:Fig2b}
\end{figure*}

Next, we acquire dephasing time maps of the sample. By the procedure described in the experimental methods section, we extract dephasing times for every pixel of the image and plot the resulting dephasing time maps in Fig.\,\ref{fig:Fig2b}(a). Before fitting a uni-exponential decay, we subtract a background determined from an image acquired after the FWM response of the sample has fully decayed. The measured sample response is commonly modeled by convolving an exponential decay fit function with the instrument response function \cite{Kasprzak_LW1,Kasprzak_LW2} to account for the finite pulse duration. However, finite pulse effects in nonlinear multi-pulse experiments are more complicated \cite{FinitePulseEffect}. Therefore, here we only fit the exponential decay for sample points with $t'\geq40$\,fs, such that $t=\tau \geq20$\,fs, while also convolving the fit function with the Gaussian instrument response function.
Fig.\,\ref{fig:Fig2b}(a) shows a joint representation of FWM intensity and dephasing times. Here, the hue level signifies the dephasing times while the saturation signifies the FWM intensity.
The high FWM intensity area of the sample shows approximately homogeneous dephasing times across the sample, except for the bottom part, which shows faster dephasing. However, several low FWM areas on the sample show a significantly increased dephasing time. 
To understand the physical reasoning behind such behavior, Fig.\,\ref{fig:Fig2b}(b) shows three exemplary dephasing curves for the sample points marked with blue, red, and green triangles in Fig.\,\ref{fig:Fig2b}(a). The red curve clearly shows a faster decay than the blue curve, establishing that the decreased dephasing time for the bottom of the sample is not an artifact of the fits but is instead reflected in the data. In contrast, the green curve initially rises before peaking at 70\,fs and decaying for larger photon echo time values. For all three curves, the fits agree well with the data in the range of 40\,fs-150\,fs for the photon-echo times, while early times show a deviating behavior due to both finite pulse and time-ordering effects during pulse overlap \cite{Jonas_timeordering}.

\begin{figure*}[b]
\centering\includegraphics[width=0.7\textwidth]{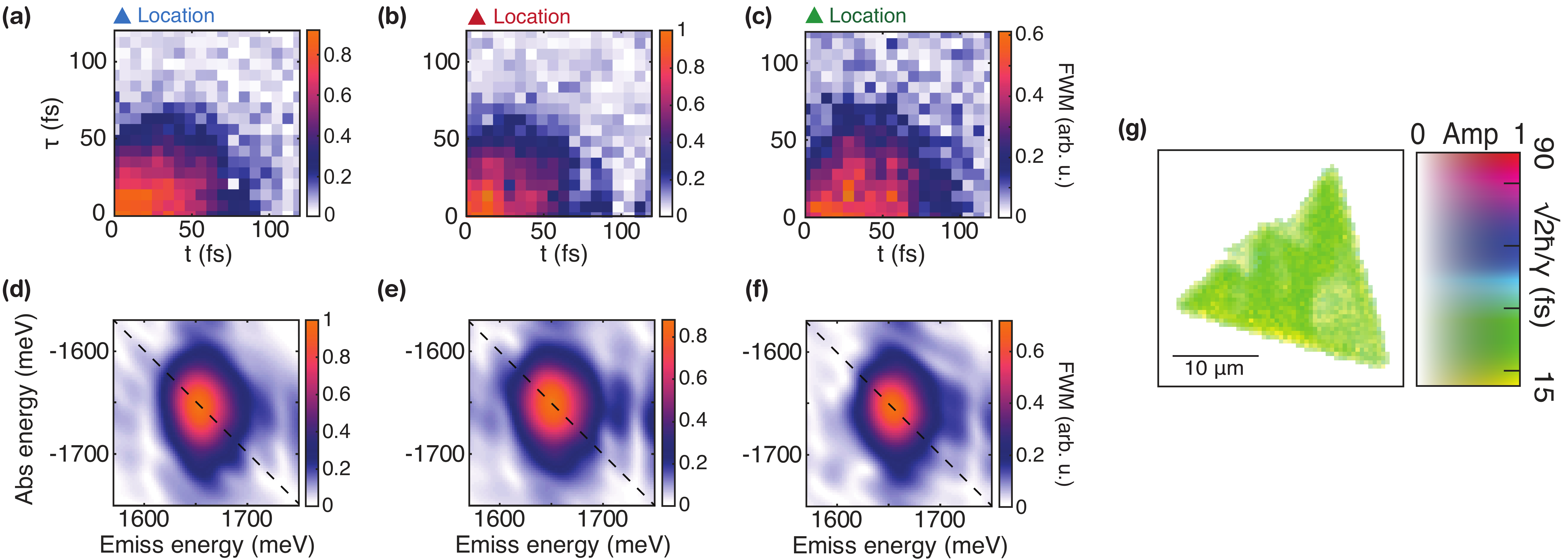}
\caption{\textbf{(a-c)} Two-dimensional time-domain signal for the three locations marked in Fig.\,\ref{fig:Fig2}(c), obtained by scanning the $\tau$- and $t$-delay. \textbf{(d-f)} MDCS spectra for the three locations marked in Fig.\,\ref{fig:Fig2b}(a), obtained by Fourier transforming the time-domain signals displayed in (a-c). \textbf{(g)} Joint map of dephasing time $T_2$ map of the sample obtained by fitting the linewidths of the MDCIS spectrum with the procedure outlined in \cite{Fits} and FWM strength.}
\label{fig:Fig3}
\end{figure*}

A better understanding of the physical behavior that causes the non-trivial temporal behavior for the sample region marked by the green triangle in Fig.\,\ref{fig:Fig2b}(a) can be gained by taking a full MDCIS scan at $T$=25\,fs. Exemplary two-dimensional time-domain signals and MDCS spectra at the three sample spots marked in Fig.\,\ref{fig:Fig2b}(a) are plotted in Fig.\,\ref{fig:Fig3}(a-c) and (d-f), respectively. The time-domain photon echo signal and corresponding spectra for Fig.\,\ref{fig:Fig3}(a,b) and (d,e) show similar behavior, with the echo in Fig.\,\ref{fig:Fig3}(b) showing a slightly faster decay, and hence a slightly increased homogeneous linewidth in the spectrum presented in Fig.\,\ref{fig:Fig3}(e). In contrast to these two sample spots, the sample spot marked with a green rectangle whose MDCIS data is plotted in Fig.\,\ref{fig:Fig3}(c,f) shows a photon-echo that is delayed along $t$.
The delayed sample response along $t$ has been observed previously \cite{Wegener_ManyBody,Wang_ManyBody,Shacklette_ManyBody} and can be explained by increased many-body effects in this sample area.
The shift in the time-domain explains the dephasing curves' shape in Fig.\,\ref{fig:Fig2b}(b). While the extracted dephasing times $T_2$ at those sample spots are unreliable, our rapid technique still unquivocally identifies these areas, and, using MDCIS, these areas can be studied further.

To prove that the dephasing times extracted for the higher FWM signal sample areas can be reliably determined by scanning along the diagonal of the photon echo, we fit the linewidths of the MDCIS measurements using the procedure outlined in \cite{Fits}. We plot the $T_2$ times determined via $T_2 = \sqrt{2} \hbar/\gamma$ from the MDCIS measurements in Fig.\,\ref{fig:Fig3}(g).
Firstly, the qualitative changes of the linewidths across the sample extracted from the two methods agree well for the high FWM signal areas. The good qualitative agreement is especially evident for the bottom area of the sample which shows reduced dephasing times in both measurements. Furthermore,
reasonable quantitative agreement between the dephasing times extracted from the rapid dephasing curves and the MDCIS linewidths can be observed. Over large areas of the sample, the dephasing times lie within 15\,fs of each other. The systematically lower dephasing time for the MDCIS measurement can be explained by finite pulse and time-ordering effects that play a significant role in these measurement because $T_2 \approx T_{\mathrm{pulse}}$. While early delay-times where those effects dominate can be filtered out efficiently in the time-domain fits, the same treatment is not as straightforward in the frequency domain. An in-depth discussion of these effects can be found in the literature \cite{FinitePulseEffect,Jonas_timeordering} and the supplementary information \cite{Supplement}. For $T_2 > T_{\mathrm{pulse}}$, these effects become negligible.

\begin{figure*}[t]
\centering\includegraphics[width=0.7\textwidth]{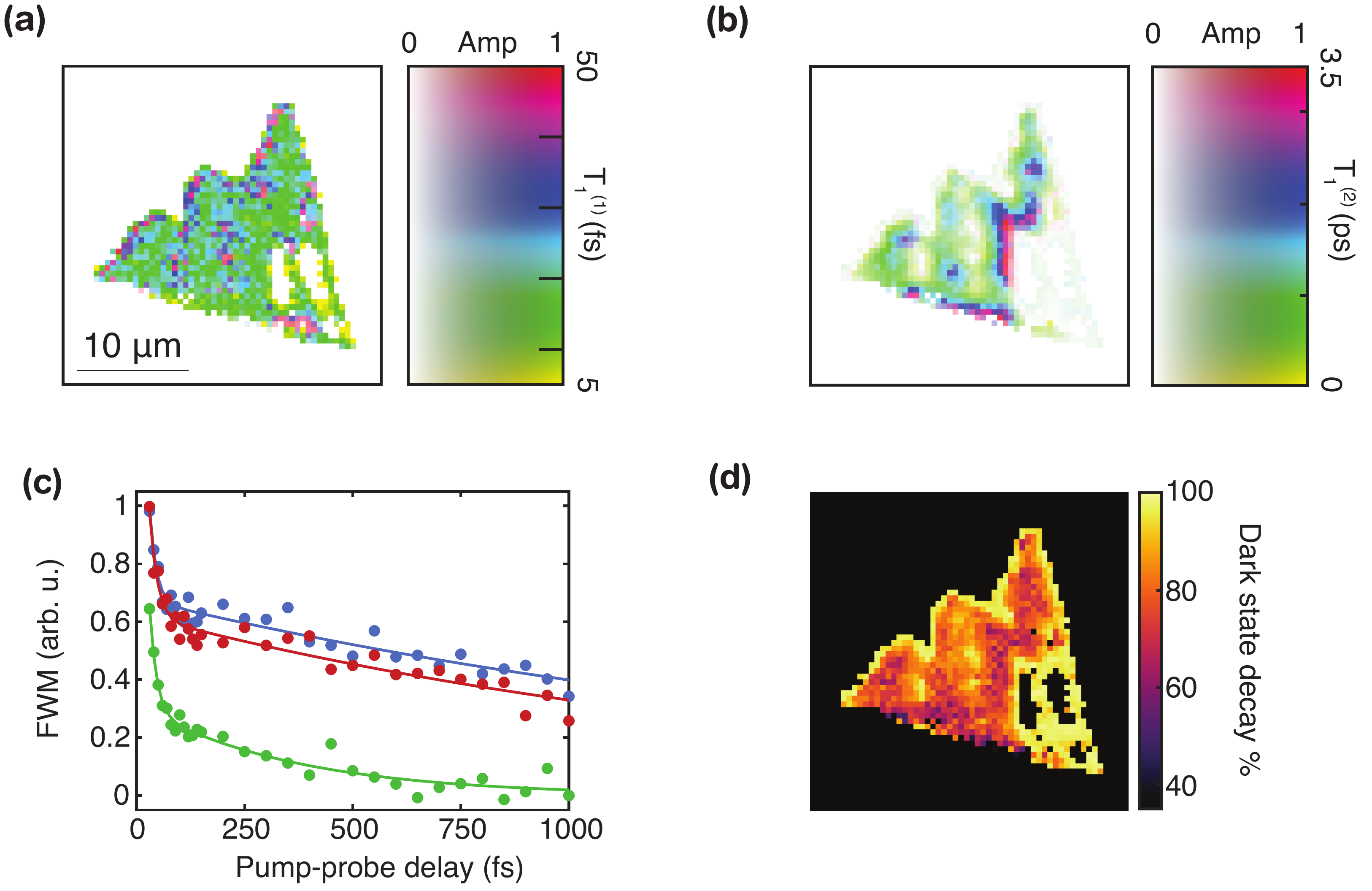}
\caption{\textbf{(a)} FWM amplitude decay curves at select sample spots marked with colored rectangles in Fig.\,\ref{fig:Fig2b}(a). Bi-exponential fits are plotted as solid lines. \textbf{(b)} Joint representation of fast fit amplitude and decay time ${T_1}^{(1)}$, obtained from the bi-exponential fit to the decay curves. \textbf{(c)} Joint representation of slow fit amplitude and decay time ${T_1}^{(2)}$, obtained from the bi-exponential fit to the decay curves. \textbf{(d)} Spatial map of the percentage of the exciton population that decays into dark states, obtained by normalizing the fit amplitude of the fast decay by the sum of slow and fast fit amplitudes.}
\label{fig:Fig4}
\end{figure*}

An additional modality of our nonlinear microscope is the ability to take rapid FWM decay images, characterizing the exciton population lifetime, by varying the $T$ delay.
We apply a bi-exponential fit to the data to capture the full temporal decay dynamics of the WSe\textsubscript{2} flake which displays rapid, sub-50\,fs decay, followed by a slower decay on the order of few picoseconds.
Similar to the dephasing time maps, we exclude sample points with $T\leq$20\,fs to avoid the influence of finite pulse effects and consider the finite pulse duration by convolving the fit function with the instrument response function. The fast decay components' fitted amplitude and decay time are plotted in the joint representation introduced earlier in Fig.\,\ref{fig:Fig2b}(a). Across the sample, the intensity of the first decay component is relatively homogeneous, while the decay time fluctuates mainly between 10-25\,fs, approximately twice as fast as the dephasing time. These observations suggest a radiatively limited dephasing time via the relation $T_2 = 2 T_1$. Although we observe a factor between 1.5-2 in our measurements, this deviation can be explained with the dephasing and decay times approaching the temporal resolution of the experimental setup. The second decay component shows a more distinct spatial profile, matching the spatial intensity profile of the FWM plotted in Fig.\,\ref{fig:Fig2}(c). The bottom and center of the sample display a longer decay time on the order of 2-3\,ps, while the rest of the sample shows decay times around 1\,ps. This behavior is further corroborated by the exemplary decay curves for the three sample spots marked in Fig.\,\ref{fig:Fig2b}(a), plotted in Fig.\,\ref{fig:Fig4}(c). The rapid sub-50\,fs decay component shows comparable amplitudes (the FWM dropping by approximately 0.4 during the first 100\,fs) for all three sample points. However, the green curve has a remaining amplitude of 0.2, while both blue and green curve have a remaining amplitude of 0.6, showing the stark contrast in amplitude for the slow decay.

Bi-exponential decays for monolayer TMDs have been observed in the past, with the fast component being attributed to the population relaxation of bright excitons and the second decay component attributed to additional states such as dark states or localized states (in the following: dark states) beyond a simple two-level system \cite{Galan_Wse2}, with the dark state constituting the ground state of the exciton in WSe\textsubscript{2} \cite{WSe2_Theory}. In this case, the fast decay is caused by decay into the dark states. After exciton populations of bright and dark states equilibrate, excitons tunneling back from dark into bright states and subsequent radiative decay of the bright excitons causes the slow decay of the signal \cite{RateEquations}. Hence, by dividing the amplitude of the fast decay by the sum of the two fit amplitudes we can further extract which percentage of the exciton population decays into dark states and observe the spatial variation of this quantity across the sample. A spatial map of this quantity is plotted in Fig.\,\ref{fig:Fig4}(d). The stronger FWM signal regions of the sample on average show a lower percentage of exciton population decaying into dark states. Among these regions, the bottom of the sample stands out with only 40\% of excitons decaying into dark states, while the center area shows approximately 60-70\% of excitons decaying into dark states. Low signal areas of the FWM show more than 90\% of excitons decaying into dark states. This corroborates earlier findings from FWM strength and dephasing maps: Low FWM signal areas of the sample, which have been identified by both their FWM strength as well as dephasing and population maps to be distinct in their physical properties from other regions of the sample also have more available dark states. Given that the exciton ground state in WSe\textsubscript{2} is a dark state, the higher number of dark states can explain the lowered measured FWM signal.

The ability to observe these physical changes across the sample is an important distinction in capabilities between ultrafast FWM imaging and other techniques presented here - resonant micro-reflectance, white light microscopy, and even the static FWM image (taken at fixed $\tau-, T-,$ and $t$-delays): While static FWM imaging at least provides an identification of areas with stronger and weaker FWM (which is a measure of underlying changes in material parameters that FWM is more sensitive two than linear imaging techniques), dephasing and decay imaging show which physical properties of the sample are altered and where. These insights into the sample changes can be incorporated into the manufacturing process: either for sample repair (e.g., through chemical processing), or within a feedback loop used to adjust the fabrication process.

\begin{figure*}[t]
\centering\includegraphics[width=0.8\textwidth]{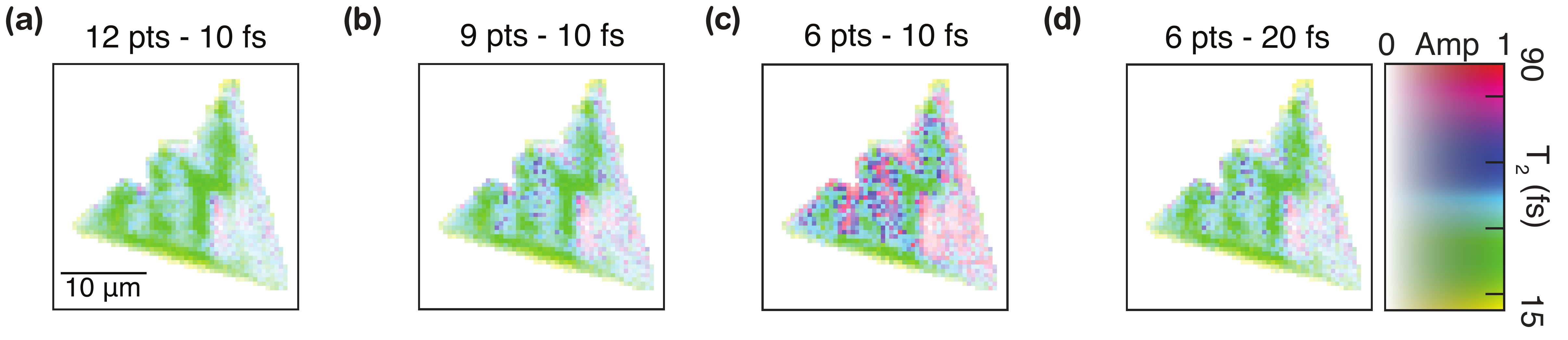}
\caption{\textbf{(a)} Dephasing time map obtained by considering all 12 time-data points per pixel between 40\,fs and 150\,fs, spaced by 10\,fs. \textbf{(b)} Dephasing time map obtained by disregarding the last three data points. \textbf{(c)} Dephasing time map obtained by disregarding the last six data points. \textbf{(d)} Dephasing time maps obtained by considering 6 data points between 40\,fs and 150\,fs, spaced by 20\,fs.}
\label{fig:Fig5}
\end{figure*}

Lastly, we turn the discussion on how to accelerate and improve the presented material characterization techniques further. In an ideal, noise-free scenario, two sample points are sufficient to determine the fit parameters for a uni-exponential decay. However, as experimental noise increases, a higher number of points increases the reliability of the extracted fit parameters. The increased noise is evident in Fig.\,\ref{fig:Fig5}, where we plot the dephasing time maps for a different selection of data points: Fig.\,\ref{fig:Fig5}(a) shows the dephasing time map obtained by considering all 12 time data points (per pixel) between 40\,fs and 150\,fs, equidistantly spaced by 10\,fs, Fig.\,\ref{fig:Fig5}(b) shows the dephasing time map obtained by omitting the last three data points, Fig.\,\ref{fig:Fig5}(c) shows the dephasing time map obtained by omitting the last six data points. As more data points get omitted, the noise for the dephasing time map visibly increases while also showing a systematic shift towards higher dephasing times. However, while using the same amount of data points as in Fig.\,\ref{fig:Fig5}(c), Fig.\,\ref{fig:Fig5}(d) uses every second data point between 40\,fs and 150\,fs, showing a significantly improved noise, effectively comparable to the dephasing time map in Fig.\,\ref{fig:Fig5}(a) which takes twice as long to acquire. The absolute values for the dephasing times in Fig.\,\ref{fig:Fig5}(a) and (d) also agree.

\begin{figure*}[t]
\centering\includegraphics[width=0.8\textwidth]{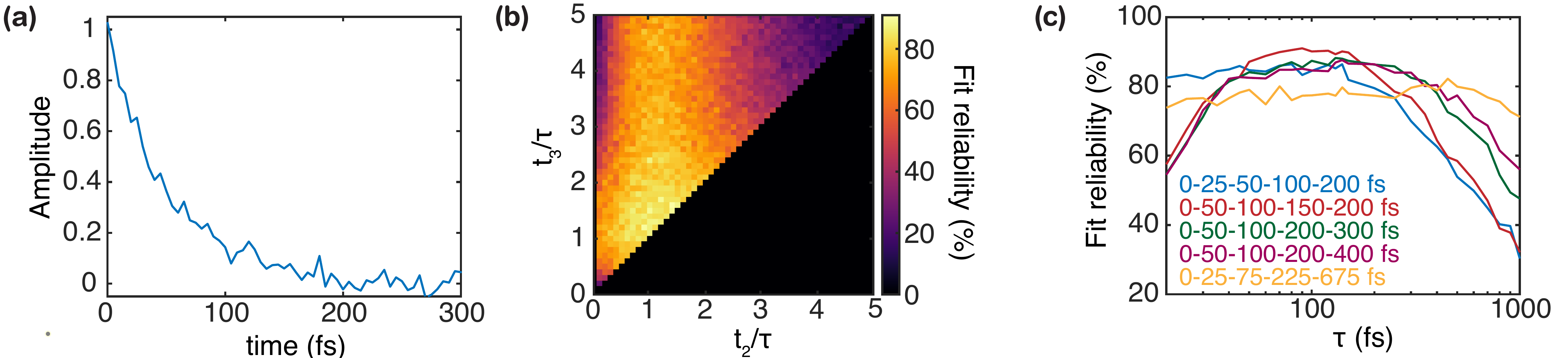}
\caption{\textbf{(a)} Uni-exponential amplitude decay with a decay time $\tau=50$\,fs. We add average white Gaussian noise with an SNR of 10 for the amplitude, corresponding to -20\,dB. \textbf{(b)} Simulated fit reliability, defined as a fitted decay constant within $\pm$10\% of the true value, for 200 iterations with randomized noise as a function of sampling point spacing for the second ($t_2$) and third ($t_3$) sample point. The first sampling point is fixed to time-zero. \textbf{(c)} Simulated fit reliability as a function of time constant for five sampling points with varying spacing.}
\label{fig:Fig6}
\end{figure*}

These observations spark the question which distribution of sample points yields the highest robustness to inevitable noise sources in the FWM measurement. To answer this question, we simulate a uni-exponential amplitude decay curve with added average white Gaussian noise with an SNR of 10 for the amplitude (-20\,dB), exemplarily plotted in Fig.\,\ref{fig:Fig6}(a). Comparing this curve to Fig.\,\ref{fig:Fig2}(f) and Fig.\,\ref{fig:Fig4}(a), we determine that this constitutes a reasonable approximation for the measurement noise level. We first simulate the scenario of three sampling points. While this is one more sampling point than necessary, this allows us to see if/how increasing the number of sampling points helps the reliability of the extracted decay times in the presence of noise.
We fix the first sampling point to time-zero and simulate a grid of sample points $t_{2,3}$ between zero and five times the decay time $\tau$ in steps of $\tau/10$. Due to symmetry conditions, we also enforce $t_3 > t_2$. For each grid point, we run 200 iterations with randomized average white Gaussian noise, all with a -20\,dB noise floor. We fit each curve and consider the fit reliable if it reconstructs the decay time $\tau=50$\,fs within $\pm$10\%. The result of the simulation is plotted in Fig.\,\ref{fig:Fig6}(b). The vertical stripe structure indicates that a reliable decay constant can almost always be extracted as long as the second sampling point is close to the decay time $\tau$. To corroborate this observation, for low values of $t_2$, a higher fit reliability can be observed for values of $t_3$ close to 1. Despite the vertical stripe structure, choosing $t_3$ in areas of meaningful signal within 2-3\,$\tau$ yields a higher reliability of the fit. In this case, the third data point contributes meaningful information to the fit instead of being close to zero, a value all exponential fits eventually approach.

However, this model considers apriori knowledge of the decay times, which is not always given. Moreover, decay times will inherently vary across the sample. Another area of interest is hence which range of decay constants can be reliably retrieved with a particular spacing of points. In the three sample point case, the answer is rather straight-forward: Setting the second sampling point to the lowest expected decay time while setting the third sample point to the largest expected decay time will yield the largest range of possible decay times. However, there is a potential decrease in reliability for intermediate values, depending on the range of decay times explored. In this case, acquiring more sample points, compromising speed for reliability to a certain extent, can be advantageous. We simulate the reliability of fits for a range of decay times from 20\,fs to 1000\,fs with five sampling points and plot the results in Fig.\,\ref{fig:Fig6}(c).
Here, we choose five different spacings: A non-equidistant spacing of sampling points where the distance between each sampling points doubles, once with an initial 25\,fs spacing (blue curve) and an initial 50\,fs spacing (purple curve), a non-equidistant spacing of sampling points where the distance is tripled, with an initial spacing of 25\,fs (yellow curve). We also simulate a scenario with an equidistant spacing of 50\,fs (red curve) and a curve with mixed spacing (green curve). We simulate a thousand fits with the same -20\,dB average white Gaussian noise for each decay time.

From Fig.\,\ref{fig:Fig6}(c), we observe that non-equidistant spacing is preferable in most scenarios. While the non-equidistant curves show a lower peak fit reliability (80\% for the blue/green/purple curve, 75\% for the yellow curve, instead of 90\% for the equidistant red curve), the range of decay times they span is significantly higher. The blue curve shows fit reliabilities above 80\% down to the smallest decay times simulated. In comparison, the purple curve shows fit reliabilities above 50\% for decay times up to 1000\,fs and down to 20\,fs. Non-equidistant spacing is also not limited to doubling the distance between the sampling points. From the yellow curve, which uses a tripled spacing between sampling points, we observe that the range of reliably extracted decay times can be further extended, with fit reliabilities over 70\% for the entire 20-1000\,fs range. However, the peak fit reliability is further decreased in this case. The sample point spacing is hence a powerful tool that can be adjusted as needed for fit reliability and extracted decay time range changes.

\section{Conclusions and Outlook}

This work presents an approach to rapid multiplex ultrafast nonlinear imaging of advanced materials based on FWM generated by three pulses. Using three pulses to generate the FWM allows us to track FWM intensity, dephasing times, and exciton lifetimes across a CVD-grown monolayer of WSe\textsubscript{2}. We show that a single FWM image alone allows to distinguish areas that show distinct exciton dephasing, population decay times, and distribution of dark states. The access to the variation of these parameters across the sample gives the ultrafast FWM modality presented in this work a selectivity that can be employed as a feedback mechanism in a fabrication setting.
FWM strength, dephasing, and decay times are material parameters reveal more about the quality of the sample than current techniques such as white light microscopy, micro-reflectance/transmission, or even photoluminescence \cite{JCP}. Furthermore, we demonstrate how to extract these parameters without compromising on acquisition speed by performing all measurements in the time-domain using few sampling points. We further show via simulations that non-equidistant spacing of the time-domain sampling points allows for the retrieval of a more extensive range of decay times at only a small cost of reduced fit reliability.
Given the broad usability of four-wave-mixing-based ultrafast imaging from two-dimensional quantum materials \cite{JCP} and defects in graphene \cite{Defects_Quantum} to distinguishing benign from malignant melanoma \cite{PumpProbe_Melanoma}, these method advancements constitute an important step forward toward material inspection in both research and industrial settings, and potentially life-science and medical imaging. 

\subsection*{Funding}

The research at the University of Michigan was supported by NSF Grant No. 2016356 and DOE grant DE-SC0022179. Development of the ultrafast nonlinear microscope at MONSTR Sense Technologies was supported by NSF Grant No. 2015068. R.U. acknowledges funding by the Max-Planck Society.

\subsection*{Acknowledgments}
We thank Adam Alfrey for helpful discussion on TMD material growth. 

\subsection*{Disclosures}
E.W.M. and S.T.C. are co-founders of MONSTR Sense Technologies, LLC, which sells ultrafast spectrometers and laser-scanning microscopes:

\noindent TLP (P), STC (I, P), EWM (I, E, P).

\subsection*{Data Availability Statement}
 Data underlying the results presented in this paper are not publicly available at this time but may be obtained from the authors upon reasonable request.

\subsection*{Supplemental document}
See Supplement for supporting content on pulse compression and characterization, a discussion of finite pulse effects, and population decay curves for early pump-probe delay times.

\bibliography{sample}

\end{document}